\documentclass[modern,tighten]{aastex631}
\makeatletter
\let\frontmatter@title@above=\relax
\makeatother
\usepackage{amsmath}
\usepackage[normalem]{ulem}

\newcommand{\eb}{\begin{equation}}
\newcommand{\ee}{\end{equation}}

\newcommand{\uasyr}{$\mu$as yr$^{-1}$}

\usepackage{subfigure,graphicx}
\usepackage{url}
\usepackage{color}
\usepackage{booktabs}
\usepackage{changepage}
\usepackage[flushleft]{threeparttable}
\usepackage{tabularx}
\definecolor{darkgray}{gray}{0.4}
\definecolor{patinared}{rgb}{.72,.10,0}
\definecolor{patinablue}{rgb}{0,.20,.65} 
\definecolor{orange}{rgb}{1,0.5,0}
\definecolor{rkka}{RGB}{219,66,32}
\definecolor{151}{rgb}{0.1,0.5,0.1}

\hypersetup{
  colorlinks=True,      
  urlcolor=darkgray,    
  linkcolor=patinared,  
  citecolor=patinablue, 
}

\shorttitle{Mapping the Universe as a Bianchi I Cosmology}
\shortauthors{Makarov et al.}

\begin{document}

\title{Mapping the Universe as a Bianchi I cosmology with Gaia data}

\correspondingauthor{Valeri Makarov} \email{valeri.makarov@gmail.com}
\author[0000-0003-2336-7887]{Valeri V. Makarov}
\affiliation{U.S. Naval Observatory, 3450 Massachusetts Ave NW, Washington, 20392-5420, DC, USA}
\author{Asta~Heinesen}
\affiliation{Department of Physics and Astronomy, Queen Mary University of London, UK}
\affiliation{Niels Bohr Institute, Blegdamsvej 17, DK-2100 Copenhagen, Denmark}
\author{Thomas Sch\"ucker}
\affiliation{Aix Marseille Univ, Universit\'e de Toulon, CNRS, CPT,  Marseille, France}


\begin{abstract}
Measurements of tangential drifts of distant quasars and galactic nuclei on the celestial sphere provide a novel and independent method of testing cosmological hypotheses. 
In this work, we employ an axisymmetric Bianchi I model as a relatively simple phenomenological model that is useful for quantifying departures from the cosmological principle.
Using a quality-filtered sample of 1.2 million proper motion vectors of distant quasars from Gaia Data Release 3, we perform global fits of the position drift fields with vector spherical harmonics (VSH) to second degree for five non-overlapping subsets of the sources with redshifts from 0.5 to 3, and assess the ability of the Bianchi I model to describe the signal. 
We theoretically demonstrate that an axisymmetric Bianchi I model produces a signal that can be described as a single quadrupole VSH term with an eigendirection which is aligned with the axis of maximum expansion anisotropy. 
We estimate this preferred direction from the Gaia data and the VSH fit, and perform point-estimates of the amplitude of the signal as a function of redshift. Although a significant quadrupole signal is detected in each bin, the increase of the amplitude of the signal with redshift predicted by the Bianchi I model is not confidently confirmed. The estimated value of the local expansion shear is higher than expected.
Possible advances in describing the kinematic patterns of a high-redshift Universe with more complex cosmologies accommodating time-dependent anisotropy and rotation are discussed.
\end{abstract}

\section{Introduction} \label{int.sec} 
The drifts of sources at cosmological distances across our sky are conventionally used to determine the local acceleration of the Solar System relative to the cosmological rest frame. 
However, drifts in sky positions of cosmic sources hold a wealth of additional information on physics at a hierarchy of scales -- from intergalactic kinematics to large-scale cosmological expansion -- which can be uncovered from the data. 
For example, position drift can be used to uncover peculiar motions and bulk flows \citep{Marcori:2018cwn},  the gravitational wave background \citep{gwenergydensity}, intrinsic cosmic shear \citep{Marcori:2018cwn}, vorticity \citep{2012ApJ...755...58N,Amendola:2013bga}, as well as Ricci and Weyl curvature of the space-time \citep{Heinesen:2024npe}. 

The cosmological imprints on the position drift signal are expected to be subdominant to the local acceleration of the Solar System within our Galaxy and Local Group. 
Recent analyses, however, showed that the low multipoles in the position drifts (proper motions) of extragalactic quasi stellar objects in the Gaia celestial reference frame catalog (data release 3) are not exclusively dominated by the constant irrotational dipole signal expected from our local acceleration: there is moderate evidence (2-3$\sigma$) of changes of the dipole amplitude with redshift by as much as $\sim 50\%$, and also a highly significant quadrupole signature that persists to redshifts of $\sim 6$ \citep{2025NatAs...9.1396M,2025arXiv250802810T}. 
These features could be ascribed to unaccounted for systematics in the Gaia instrumentation or to signatures of new cosmological physics. 
If caused by systematic effects, this may put into question the robustness of the independent detection of the Solar System acceleration by Gaia itself, as the identified  systematics would in this case be comparable to the size of the reported acceleration signal, and it is thus important to investigate this direction carefully. 

In this paper, we investigate the potential of new cosmological physics in the data. 
We do this by considering an axisymmetric Bianchi I universe model as a simple model with cosmic shear, which naturally gives rise to a quadrupole in the observed proper motions of extragalactic sources. 
Since the Bianchi I universe model is globally anisotropic, our analysis may also be viewed as providing a test of the cosmological principle. 
While the Bianchi I universe can be viewed as a simplistic toy model, it provides a physical framework for interpreting the coherent drifts of cosmic sources across our sky that are visible in the low multipoles of the Gaia Celestial Reference Frame (CRF) proper motion field. The task of empirical mapping of the Universe can be approached from the bottom up, i.e., by starting with simpler and cleaner concepts and moving up the hierarchy of more complex models if we find compelling inadequacies and tensions with the available data.

\section{Cleaned Gaia DR3 sample of CRF quasars}
\label{data.sec}
A detailed description of the data sample used in this study can be found in \citep{2025NatAs...9.1396M}. The initial dataset is based on the Gaia CRF catalog (data release 3) \citep{2023A&A...674A...1G, 2022A&A...667A.148G} cross-matched with the main Gaia catalog to extract additional information needed for proper motion computations. The initial collection includes 1.56 million sources that are predominantly distant AGNs and quasars. The information in the joint table is complemented with machine-learning based redshift estimates and the observed spectroscopic redshifts from the SDSS DR16 catalog \citep{2020ApJS..249....3A}. 
The estimated redshifts were generated using a trained neural network algorithm and several classifier parameters, including the optical and mid-infrared colors, and Gaia metadata parameters. 
In the light of a recent analysis of hidden dependencies of the Gaia CRF proper motion systematics on a much smaller collection of radio-optical reference objects, an additional cut was applied based on the parameter representing the number of astrometric unknowns in the Gaia solution. The majority of the CRF sources were solved in the regular option with 5 astrometric unknowns per object (position and proper motion components, parallax), while the rest were treated in a special 6-parameter solution including a pseudocolor. The latter solutions were shown to have a significantly different systematics because the calibration procedures were more complex and less stable \citep{2021A&A...649A...2L}. All the sources with 6-parameter solutions were discarded for this analysis resulting in a sample of 1.2 million sources. 

Despite all the effort taken to eliminate interlopers and look-alikes of quasars, the Gaia CRF catalog still includes a small fraction of relatively nearby objects such as ultraluminous stars in nearby galaxies. A handful of specific examples has been identified \citep{2024ApJS..274...27M}. Even at a low rate of occurrence, such objects can affect the vector spherical harmonics (VSH) decomposition fit of the overall proper motion field in a weighted least-squares adjustment due to their relative brightness and small formal errors. To mitigate this risk, a filter on the standardized proper motion magnitude was implemented in this study. If $\boldsymbol{\mu}_i$ is the given proper motion vector in the catalog and $\boldsymbol{C}_{ii}$ is its formal covariance matrix, the quadratic form $\boldsymbol{\mu}_i^T \boldsymbol{C}^{-1}_{ii} \boldsymbol{\mu}_i$ is expected to follow the $\chi^2(2)$ distribution provided the vectors are drawn from a bivariate normal distribution with zero mean. Their square roots are Rayleigh(1)-distributed, so that 5.6\% of objects should have magnitudes exceeding 2.4. The actual sample includes 7.7\% of such outliers. The excess 2.1\% may reflect slightly underestimated formal errors, or may include contaminants. To be on the safe side, we removed all sources with standardized magnitudes above 2.4 resulting in a working sample of 1,105,945 objects.

We note that the data sample includes only objects with redshifts (spectroscopic or estimated) above 0.5. The relatively closer AGNs and quasars have been shown to possess discordant astrometric characteristics in the Gaia catalog \citep[e.g., ][and references therein]{zhang2026astrometricpropertiesreferenceframe}. These sources are often associated with host galaxies, whose surface brightness is non-negligible and the images are not consistent with the assumed point-source morphology \citep{2012MmSAI..83..952M}.


\section{Computation of empirical proper motion fields}


Tangential vector spherical harmonic 
(VSH) functions can be viewed as radial projections of the 3D vector field derived from the scalar spherical harmonics $Y_{p,q}$, where $p$ is the degree and $q$ is the order of the spherical harmonic, on the unit sphere. 
The tangential component of VSH are obtained as gradient and curl operators, $\boldsymbol{E}_{p,q}=\boldsymbol{\nabla} Y_{p,q}$ and $\boldsymbol{M}_{p,q}=\boldsymbol{\hat{n}}\times \boldsymbol{\nabla} Y_{p,q}$, where $\boldsymbol{\hat{n}}$ is then radial unit vector. We refer to $\boldsymbol{E}_{p,q}$ and $\boldsymbol{M}_{p,q}$ as the electric and magnetic harmonics, respectively. Any continuous and everywhere differentiable velocity field $\boldsymbol{\mu}$ on the unit sphere can be expressed as
\eb 
\boldsymbol{\mu}(l,b)=\sum_{\{k,p,q\}}\left(a_{\{k,p,q\}}^{(E)}\boldsymbol{E}_{\{k,p,q\}}(l,b)+a_{\{k,p,q\}}^{(M)}\boldsymbol{M}_{\{k,p,q\}}(l,b)\right) . 
\label{vsh.eq}
\ee 
The indices take values $p=1,2,\ldots,\infty$, $q=1,\ldots, p$, $k=0,1,2$. The kind parameter, $k$, has been introduced to keep track of real and imaginary parts of the harmonics, where $k=1$ label the real part and $k=2$ the imaginary part. Kind $k=0$ is used separately for the (real) zonal VSH functions, which depend only on the second coordinate $b$, and therefore have $q=0$. The complete set of VSH functions up to degree $P$ includes $2P(P+2)$ terms. 
Our choice of galactic coordinate system $(l,b)$ is such that we have a Galactic $X$ axis pointing toward the Galactic center. 

In our analysis, we estimate the 16 VSH coefficients that characterize $\boldsymbol{\mu}(l,b)$ up to degree $P=2$. 
It will be shown in Section~\ref{theo.sec} that the series is sufficient to quantify the effects of a Bianchi I cosmology and other signals related to the observer's motion in space. 
The first- and second-degree VSH functions are listed in Table \ref{vshlist.tab} with their names given in the tuple form as in \citep{2025NatAs...9.1396M}: 
$\boldsymbol{E}_{\{k,p,q\}} = \{{\rm ele}, k, p, q\}$ and $\boldsymbol{M}_{\{k,p,q\}} = \{{\rm mag}, k, p, q\}$.

In the framework of generalized least squares method, we compute the fitting VSH coefficients minimizing the
$\chi^2$ statistic\footnote{The technical implementation, employing a least-squares adjustment is described in \citep[][supplementary materials]{2025NatAs...9.1396M}.}  
\eb 
\chi^2 = \sum_{i=1}^{n} \left(\boldsymbol{\mu}_i - \boldsymbol{\mu}(l,b) \right)^T \boldsymbol{C}^{-1}_{ ii} \left(\boldsymbol{\mu}_i - \boldsymbol{\mu}(l,b) \right) ,
\ee 
where $\boldsymbol{\mu}_{i}$ is the measured drift of the i'th source, and $\boldsymbol{\mu}(l,b)$ is the corresponding VSH prediction for the source given by \eqref{vsh.eq} as truncated at degree $P=2$. 
The $\boldsymbol{C}_{ii}$ is the $2 \times 2$ covariance matrix for the i'th source, and $n$ is the number of sources in the sample.  
The observational data (Gaia DR3 proper motions $\boldsymbol{\mu}_i$ and their formal covariance matrices $\boldsymbol{C}_{ii}$) are transformed from the original equatorial system to the Galactic coordinate system employed in this analysis, using standard transformation procedures \citep{1997ESASP1200.....E}. 



\begin{deluxetable*}{L|L|C}
\tablecaption{VSH functions to degree 2.}
 \label{vshlist.tab}
\tablehead{
\colhead{j} & \colhead{\text{name}} & \colhead{\text{VSH function}}
}
\startdata
 1 & \{\text{mag},0,1,0\} & \left[-\frac{1}{2} \sqrt{\frac{3}{\pi }} \cos (b),\;0\right] \\
 2 & \{\text{mag},1,1,1\} & \left[\frac{1}{2} \sqrt{\frac{3}{2 \pi }} \sin (b) \cos(l),\;-\frac{1}{2} \sqrt{\frac{3}{2 \pi }} \sin (l)\right] \\
 3 & \{\text{mag},2,1,1\} & \left[\frac{1}{2} \sqrt{\frac{3}{2 \pi }} \sin (b) \sin(l),\;\frac{1}{2} \sqrt{\frac{3}{2 \pi }} \cos (l)\right] \\
 4 & \{\text{ele},0,1,0\} & \left[0,\;-\frac{1}{2} \sqrt{\frac{3}{\pi }} \cos (b)\right] \\
 5 & \{\text{ele},1,1,1\} & \left[\frac{1}{2} \sqrt{\frac{3}{2 \pi }} \sin(l),\;\frac{1}{2} \sqrt{\frac{3}{2 \pi }} \sin (b) \cos (l)\right] \\
 6 & \{\text{ele},2,1,1\} & \left[-\frac{1}{2} \sqrt{\frac{3}{2 \pi }} \cos(l),\;\frac{1}{2} \sqrt{\frac{3}{2 \pi }} \sin (b) \sin (l)\right]\\ 
 7 & \{\text{mag},0,2,0\} & \left[\frac{3}{4} \sqrt{\frac{5}{\pi }} \sin (2b),\;0\right] \\
 8 & \{\text{mag},1,2,1\} & \left[\frac{1}{2} \sqrt{\frac{15}{2 \pi }} \cos (2 b) \cos(l),\;\frac{1}{2} \sqrt{\frac{15}{2 \pi }} \sin (b) \sin (l)\right] \\
 9 & \{\text{mag},2,2,1\} & \left[\frac{1}{2} \sqrt{\frac{15}{2 \pi }} \cos (2 b) \sin(l),\;-\frac{1}{2} \sqrt{\frac{15}{2 \pi }} \sin (b) \cos (l)\right] \\
 10 & \{\text{ele},0,2,0\} & \left[0,\;\frac{3}{4} \sqrt{\frac{5}{\pi }} \sin (2b)\right] \\
 11 & \{\text{ele},1,2,1\} & \left[-\frac{1}{2} \sqrt{\frac{15}{2 \pi }} \sin (b) \sin(l),\;\frac{1}{2} \sqrt{\frac{15}{2 \pi }} \cos (2 b) \cos (l)\right] \\
 12 & \{\text{ele},2,2,1\} & \left[\frac{1}{2} \sqrt{\frac{15}{2 \pi }} \sin (b) \cos(l),\;\frac{1}{2} \sqrt{\frac{15}{2 \pi }} \cos (2 b) \sin (l)\right] \\
 13 & \{\text{mag},1,2,2\} & \left[-\frac{1}{4} \sqrt{\frac{15}{2 \pi }} \sin (2 b) \cos (2 l),\;\frac{1}{2} \sqrt{\frac{15}{2 \pi }} \cos (b) \sin (2l)\right] \\
 14 & \{\text{mag},2,2,2\} & \left[-\frac{1}{4} \sqrt{\frac{15}{2 \pi }} \sin (2b) \sin (2l),\;-\frac{1}{2} \sqrt{\frac{15}{2 \pi }} \cos (b) \cos (2 l)\right] \\
 15 & \{\text{ele},1,2,2\} & \left[-\sqrt{\frac{15}{2 \pi }} \cos (b) \sin (l) \cos  (l),\;-\frac{1}{4} \sqrt{\frac{15}{2 \pi }} \sin (2 b) \cos (2 l)\right] \\
 16 & \{\text{ele},2,2,2\} & \left[\frac{1}{2} \sqrt{\frac{15}{2 \pi }} \cos (b) \cos(2 l),\;-\frac{1}{4} \sqrt{\frac{15}{2 \pi }} \sin (2b) \sin (2l)\right] \\
\enddata
\end{deluxetable*}

Global VSH fits involving a large number of objects allow us to boost the signal-to-noise ratio by orders of magnitude. The median magnitude of Gaia CRF proper motions is 480 \uasyr, while the effects we are trying to detect are of magnitude $\lesssim 1$ \uasyr. A positive detection thus requires $\gtrsim 10^5$ objects. This consideration drives our decision to divide the initial CRF sample into only five redshift bins. Additional experiments with larger sets of fitting functions up to degree $P=4$ were conducted to estimate the robustness of the presented results to an expanded model \citep{2025arXiv250802810T}. Although a few formally significant higher-degree terms emerge in these verification fits, they do not seem to change the quadrupole coefficients outside their formal uncertainties.

\section{Theoretically expected proper motion fields in Bianchi I cosmologies}
\label{theo.sec}

A Bianchi I universe is characterized by a generalization of the Robertson-Walker spacetime metric allowing for three different directional scale factors in the line element: 
\eb 
d\tau^2=dt^2-a^2(t)dx^2-b^2(t)dy^2-c^2(t) dz^2,
\ee 
where $\{x,y,z\}$ are the three orthogonal principal axes of the metric, and the functions $a$, $b$, and $c$ are positive at all cosmic times $t$. A Bianchi I universe is homogeneous but anisotropic. A special case is an axially symmetric universe with (for example) $a(t)=b(t)$, which has only one principal axis $z$ with a different scale. In the following analysis, the scale function $b(t)$ is absent, and $b$ denotes a spherical latitude coordinate. The three Killing vectors $\partial_x$, $\partial_y$, and $\partial_z$ correspond to translations along the principal axes. The fourth Killing vector $x\partial_y-y\partial_x$ corresponds  to a rotation around the $z$-axis. Since the metric is invariant  with respect to this rotation, the choice of the $x$- and $y$-axes becomes arbitrary.

\citet{Marcori:2018cwn} provide an in-depth analysis of drifts in tri-axial Bianchi I universes. For the axial case,
\citet{2025CQGra..42g5002S} derived specific formulae describing the expected angular displacement of distant sources of electromagnetic radiation with time, i.e. the emerging proper motion field. These equations are rendered in two spherical coordinates $\varphi$, which is the azimuth in the equatorial plane $\{x,y\}$, and $\theta$, which is the polar distance angle between the source and the $z$-axis (his Eqs. 63 and 66, respectively). It is practical to rewrite these two equations in terms of provisional astronomical coordinates $\hat l$ (longitude) and $\hat b$ (latitude): 
\eb 
\begin{split}
    \hat l&=\varphi\\
    \hat b&=\pi/2-\theta
\end{split}
\label{lb}
\ee 
defined such that $\hat b =\pm \pi/2$ degrees corresponds to the axis of symmetry,
leading to
\eb 
\begin{split}
    \delta \hat l\cos \hat b&=-\sin \hat l\, V_x\,\epsilon_d\\
    \delta \hat b&=\left[-3\sin \hat b\cos \hat b\,\left(\frac{1}{2}a_{F0}\chi_{e0}\eta'_0+\eta_e\right)-\cos \hat l\sin \hat b\,V_x+\cos \hat b\,V_z\right]\,\epsilon_d
\end{split}
\ee 
These equations are linearized to first order deviations of the scale factors $a$ and $c$ from a single Friedman scale factor $a_F$ using (small) updates $\eta$ defined via $a=: a_F\,[1-{\textstyle\frac{1}{2}}  \eta]$, 
$c=:  a_F\,[1+ \eta]$. The prime in this equation denotes the derivative of $\eta$ with respect to cosmic time $t$. This derivative itself is a function of time to be determined in equation (\ref{etat}).
 
The Friedman scale factor evaluated today (at $t_0$) is denoted by $a_{F0}$ and
$\chi_{e0}$ is the conformal time of flight of the photon from emission (e) to observation (0):
\eb 
\chi_{e0}=\int_{t_e}^{t_0}\frac{dt}{a_F}.
\ee
The common factor $\epsilon_d$ is the observer's time interval of measurement $\Delta T$ normalized by the time of flight:
\eb 
\epsilon_d=\Delta T/(a_{F0}\chi_{e0}).
\ee 

The terms linear in the Cartesian components $V_x$ and $V_z$ are generated by the peculiar velocity of the observer. These components are normalized by the speed of light. Using the invariance of the metric in the equatorial plane, the $x$-axis is chosen to coincide with the equatorial projection of the observer's velocity vector. This explains the absence of $V_y$ terms. In a more general setup where the $x$-axis is not aligned with the velocity, the differential equations of drift become
\eb 
\begin{split}
\delta \hat l\cos \hat b=(-\sin \hat l\, V_x+\cos \hat l\,V_y)\,\epsilon_d\\[1mm]
    \delta \hat b=\left[-3\sin \hat b \cos \hat b\,\left(\frac{1}{2}a_{F0}\chi_{e0}\eta'_0+\eta_e\right)-\cos \hat l\sin \hat b\,V_x-\sin \hat l\sin \hat b\,V_y+\cos \hat b\,V_z\right]\epsilon_d
    \end{split}
    \label{c.eq}
\ee 
For a direct comparison with astrometric proper motions, we set $\Delta T=1$ yr. Note that the observer's velocity and cosmological scale terms become decoupled only in the linear approximation, which uses the Taylor expansion of the drift to first order. This approximation is adequate for all practical reasons, as the higher-order terms are negligibly small. 

Substituting $\delta \hat l\cos \hat b/\Delta T\equiv \mu_{\hat l}$ and $\delta \hat b/\Delta T\equiv \mu_{\hat b}$, where $\mu$ is the emergent yearly proper motion in the tangent coordinates at point $(\hat l,\hat b)$ \citep{1997ESASP1200.....E}, one obtains the following equation for the cosmic proper motion field:
\eb 
\label{mu.eq}
\boldsymbol{\mu}=
-\sqrt{\frac{\pi}{5}}(\eta'_0+2\bar{\eta}_e)\boldsymbol{E}_{\{0,2,0\}}-2\sqrt{\frac{2\pi}{3}}\bar{V}_x\boldsymbol{E}_{\{1,1,1\}}
-2\sqrt{\frac{2\pi}{3}}\bar{V}_y\boldsymbol{E}_{\{2,1,1\}}-2\sqrt{\frac{\pi}{3}}\bar{V}_z\boldsymbol{E}_{\{0,1,0\}},
\ee 
where $\bar{V}_j$ are normalized components of reflex angular velocity, $\bar{V}_j=V_j/(a_{F0}\chi_{e0})$, and $\bar{\eta}_e=\eta_e/(a_{F0}\chi_{e0})$ is the normalized scale deviation. The velocity-related components are consistent with the well-known formulae for ``cosmic parallax" \citep{1935ZA......9..290M, 1986AZh....63..845K} for the case of zero curvature and non-vanishing cosmological constant $\Lambda$. A note about the nomenclature is proper here. The numerically larger effect related to the observer's linear motion in the cosmic frame is known in the literature on astrometry and Galactic astronomy as the ``reflex proper motion", while the term ``parallax" is commonly reserved for the annual parallax caused by the orbital motion of Earth around the Solar system barycenter. The cosmological annual parallax is more than an order of magnitude smaller (unless we live in a closed universe and measure extremely distant sources beyond the curvature radius). We will return to the prospects of measuring the cosmological reflex proper motion and the basic parameters of the Universe in Section \ref{disc.sec}.

Eq. \ref{mu.eq} tells us that the astrometric effects emerge in the considered model only in low-degree electric harmonics. Specifically, the reflex motion terms are fully represented by the three electric harmonics of first degree (dipoles), while the axial anisotropy term is described by a quadrupole electric term. There are no magnetic terms (rotations, torsions, curls) in the considered model. The separation of the velocity and anisotropy-related effects in different degrees of the VSH field should allow us to determine both phenomena, in principle. 

If we adopt Einstein's equations with a cosmological constant $\Lambda$ and comoving dust, then $a_F$ and $\eta$ are given by,
\begin{eqnarray}
a_F(t) = a_{F0}\left(\frac{\cosh\left(\sqrt{3 \Lambda}\, t\right) - 1}{\cosh\left(\sqrt{3 \Lambda}\, t_0\right)  - 1}\right)^{1/3},
\hspace{4mm} 
t_0={\textstyle\frac{2}{3}}\,\frac{\text{arc}\tanh\,\sqrt{\Omega _{\Lambda 0}}}{\sqrt{\Omega _{\Lambda 0}}}\, \frac{1}{H_{F0}}\,,
\label{scale}
\end{eqnarray}
  and
\begin{eqnarray}
\eta' = \eta'_0 \,\frac{a_{F0}^3}{a_F^3}\  
   \Rightarrow\ 
\eta(t) =
-\,{\textstyle\frac{2}{3}} \,\frac{\eta'_0}{H_{F0}} \,
\frac{\sqrt{\Omega _{\Lambda 0}}}{1-\Omega _{\Lambda 0}} 
\left( \coth\left({\textstyle\frac{1}{2}}  \sqrt{3 \Lambda}\, t\right)-
1/\sqrt{\Omega _{\Lambda 0}}
\right).
\label{etat}
\end{eqnarray}

Let us take $a_{F0} = 1$ Gyr, $\Omega _{\Lambda 0}=0.7$ and $H_{F0}= 72$ km/s/Mpc implying $1/H_{F0}=13.58$ Gyr, $t_0=13.09$ Gyr, $\chi _0=44.88$, and $1/\sqrt{\Lambda }=9.37$ Gyr.
The physical quantities of interest do not depend on $a_{F0}$ and its value is just a convenient choice of units, which attributes dimensions of time to the scale factor. 
The conformal time 
(we suppress the index $\cdot _F$ to alleviate notations)
\begin{eqnarray} \label{chi.eq}
\chi(t) := \int_0 ^t \,\frac{d \tilde t}{ a_F(\tilde t)}\,, \quad \text{implying} \quad
\chi _{e0} = \chi(t_0)-\chi (t_e)=:\chi _0-\chi _e,
\end{eqnarray}
 is by definition a dimensionless variable
and does depend on the choice of $a_{F0}$. On the other hand the product $a_{F0}\,\chi(t) $ does not depend on this choice. 

In the proper motion field, (\ref{mu.eq}), we meet the coefficent,
\begin{eqnarray}
C(t_e)&:=& 1+2\,\frac{\bar\eta_e}{\eta'_0}\, 
\nonumber\\[2mm]
&=&
1-\,\frac{4}{3}\,\,\frac{1}{H_{F0} a_{F0}\chi _{e0}(t_e)}\, 
\frac{\sqrt{\Omega _{\Lambda 0}}}{1-\Omega _{\Lambda 0}} 
\left[ \coth\left({\textstyle\frac{1}{2}}  \sqrt{3 \Lambda}\, t_e\right)-
1/\sqrt{\Omega _{\Lambda 0}}
\right] .
\end{eqnarray} 

Note that this coefficient does not depend on the anisotropy and is to be computed from the solutions of Friedman's equation given in Appendix A. We make two intermediate steps in computing $C$ from the observed redshift $z$. (The latter of course has nothing to do with the third Cartesian coordinate.)

{\it First step, conformal time of flight $\chi _{e0}$}:
Equation (\ref{conformalAge}) gives the age of the universe today in dimensionless conformal time $\chi _0$ and equation (\ref{conformalEmTime}) gives the conformal emission time $\chi _e$ of a photon arriving today with a redshift $z$. Then we compute the conformal time of flight $\chi _{e0}=\chi _0-\chi _e$.

{\it Second step, cosmic emission time $t_e$}:
 The central equation (\ref{rr2}) passes from conformal to cosmic time $t_e=t(\chi _e)$.
 
 There is also an alternative way to compute $C$:
 
 {\it First step: cosmic emission time $t_e$}:
 \begin{eqnarray}
  t_e=\,\frac{1}{\sqrt{3\Lambda }}\,{\rm arc}\cosh\left(2\,
  \frac{\Omega _{\Lambda 0}}{1-\Omega _{\Lambda 0}}\,
 \frac{1}{(z+1)^3}+1\right).
\end{eqnarray} 
 
{\it Second step, conformal time of flight $\chi _{e0}$}: use equation (\ref{uu}).

Note that computing the drift to first order does not involve the   direction dependence in the redshift, equation (53)  in \citet{2025CQGra..42g5002S}, 
\begin{eqnarray}
 z+1&=&\frac{a_{F0}}{a_{F}(t_e)}
\left[ 1+ {\textstyle\frac{1}{2}} 
\left(1-3\cos^2\theta \right)\eta (t_e)
+\sin\theta\left(\cos\varphi\, V_x+\sin\varphi \,V_y\right)+\cos\theta \,V_z \right]
\nonumber\\[1mm]
&&+\,O(\eta^2,V^2,\eta\,V). \label{zte}
\end{eqnarray} 
This simplification does not occur when computing the Lema{\^i}tre-Hubble diagram to first order in Bianchi I universes, 
\citet{Schucker:2014wca}.

 Table \ref{coef} shows the values of $C$ for the mean and boundary values of the redshift batches, which we will use later.

\begin{table}[!h]
\begin{center}
\caption{Coefficients $C$,  conformal and  cosmic emission times, and their ranges\\ $\textcolor{yellow}{.}\hspace{95mm}$ for the 5 batches of redshift}
 \label{coef}
 \vspace{5mm}
\begin{tabular}{|c|c|c|c|}
\hline
&&&\\
$z$&$t_e\ \text{in\ Gyr}$&$\chi _{e0}$&$C$
\\[2mm]\hline\hline
$0.755\pm0.255$&$6.68\,[8.19,5.55]$&$8.44\,[5.99,10.55]$&$-2.74\,[-2.11,-3.41]$
\\\hline
$1.22\pm0.21$&$4.83\,[5.55,4.25]$&$12.07\,[10.55,13.42]$&$-3.98\,[-3.41,-4.57]$
\\\hline
$1.58\pm0.15$&$3.90\,[4.25,3.60]$&$14.29\,[13.42,15.10]$&$-4.99\,[-4.57,-5.43]$
\\\hline
$1.945\pm0.215$&$3.22\,[3.60,2.91]$&$16.17\,[15.10,17.13]$&$-6.06\,[-5.43,-6.71]$
\\\hline
$2.58\pm0.42$&$2.42\,[2.91,2.05]$&$18.77\,[17.13,20.15]$&$-8.02\,[-6.71,-9.37]$
\\\hline
\end{tabular}
 \end{center}
\end{table}

\section{Results} \label{res.sec}
\begin{table}[h]
\fontsize{9.5}{11.5}\selectfont
\caption{VSH coefficients for five redshift batches of CRF quasar proper motions}\label{res.tab}%
\begin{threeparttable}
\begin{tabularx}{\linewidth}{Lr|Rc|Rc|Rc|Rc|Rc}
\toprule
\# & VSH name  & \multicolumn{1}{c}{$a$} & $\sigma_a$  & \multicolumn{1}{c}{$a$} & $\sigma_a$ & \multicolumn{1}{c}{$a$} & $\sigma_a$ & \multicolumn{1}{c}{$a$} & $\sigma_a$ & \multicolumn{1}{c}{$a$} & $\sigma_a$\\
\multicolumn{2}{c|}{} & \multicolumn{2}{c|}{batch 1} & \multicolumn{2}{c|}{batch 2} & \multicolumn{2}{c|}{batch 3} & \multicolumn{2}{c|}{batch 4} & \multicolumn{2}{c}{batch 5}\\
\multicolumn{2}{c|}{} & \multicolumn{2}{c|}{$z\in[0.50,1.01]$} & \multicolumn{2}{c|}{$z\in[1.01,1.43]$} & \multicolumn{2}{c|}{$z\in[1.43,1.73]$} & \multicolumn{2}{c|}{$z\in[1.73,2.16]$} & \multicolumn{2}{c}{$z\in[2.16,3.00]$}\\
\midrule
1 & \text{$\{$mag, 0, 1, 0$\}$} & -8.36 & 1.57 & +6.85 & 1.73 & +11.74 & 1.72 & +11.74 & 1.82 & +13.03 & 2.06 \\
 2 & \text{$\{$mag, 1, 1, 1$\}$} & +10.22 & 1.81 & -5.3 & 1.99 & -2.59 & 1.99 & -0.17 & 2.08 & +2.68 & 2.31 \\
 3 & \text{$\{$mag, 2, 1, 1$\}$} & +7.87 & 2.03 & -11.32 & 2.23 & -8.51 & 2.24 & -5.25 & 2.34 & +4.35 & 2.61 \\
 4 & \text{$\{$ele, 0, 1, 0$\}$} & +2.33 & 1.49 & -1.26 & 1.63 & -4.18 & 1.63 & +1.8 & 1.71 & +1.24 & 1.92 \\
 5 & \text{$\{$ele, 1, 1, 1$\}$} & -10.36 & 2.01 & -5.83 & 2.21 & -14.09 & 2.21 & -14.24 & 2.32 & -12.84 & 2.59 \\
 6 & \text{$\{$ele, 2, 1, 1$\}$} & -2.39 & 2.06 & +0.91 & 2.27 & +1.55 & 2.28 & -0.99 & 2.39 & -1.75 & 2.71 \\
 7 & \text{$\{$mag, 0, 2, 0$\}$} & +0.53 & 0.77 & -1.11 & 0.83 & -2.18 & 0.85 & -1.31 & 0.87 & -1.27 & 0.94 \\
 8 & \text{$\{$mag, 1, 2, 1$\}$} & -0.8 & 1.24 & -1.58 & 1.36 & -3.55 & 1.37 & -1.14 & 1.43 & -1.03 & 1.61 \\
 9 & \text{$\{$mag, 2, 2, 1$\}$} & -0.22 & 1.14 & +0.22 & 1.25 & +2.21 & 1.26 & +0.6 & 1.31 & +3.37 & 1.47 \\
 10 & \text{$\{$ele, 0, 2, 0$\}$} & -0.85 & 0.68 & +0.4 & 0.73 & -0.45 & 0.75 & -0.4 & 0.77 & -0.16 & 0.82 \\
 11 & \text{$\{$ele, 1, 2, 1$\}$} & -2.93 & 1.21 & -2.18 & 1.33 & -1.51 & 1.34 & -1.75 & 1.39 & -0.33 & 1.56 \\
 12 & \text{$\{$ele, 2, 2, 1$\}$} & -0.59 & 1.15 & +0.05 & 1.26 & -2.57 & 1.26 & +0.54 & 1.32 & -1.49 & 1.47 \\
 13 & \text{$\{$mag, 1, 2, 2$\}$} & +0.88 & 1.13 & +0.53 & 1.24 & +0.28 & 1.24 & +2. & 1.29 & +0.49 & 1.43 \\
 14 & \text{$\{$mag, 2, 2, 2$\}$} & -1.61 & 1.27 & -0.28 & 1.39 & -0.07 & 1.4 & -0.18 & 1.46 & -0.08 & 1.62 \\
 15 & \text{$\{$ele, 1, 2, 2$\}$} & +3.49 & 1.22 & +4.2 & 1.34 & +3.52 & 1.34 & +3.15 & 1.4 & +5.03 & 1.56 \\
 16 & \text{$\{$ele, 2, 2, 2$\}$} & +2.61 & 1.17 & +0.46 & 1.29 & -0.77 & 1.3 & +4.12 & 1.35 & +2. & 1.5 \\
\botrule
\end{tabularx} 
\begin{tablenotes}
  \item All VSH coefficients $a$ and their formal errors $\sigma$ are given in units \uasyr.  
\end{tablenotes}
\end{threeparttable}
\end{table}

Table \ref{res.tab} presents the fitted coefficients of the 16 VSH functions for each of the 5 redshift partitions. The 
linear regression fits were obtained from a global weighted least-squares solution including the entire dataset, as described in \citep[][ArXiv version]{2025NatAs...9.1396M}. The values are expressed in \uasyr\ as well as their formal errors (corresponding to 68\% confidence intervals) $\sigma_a$, which are the square roots of the solution covariance diagonal. The statistical significance of each term can be estimated as the formal signal to noise ratio (SNR), which is $|a|/\sigma_a$. Assuming that the distribution of errors is Gaussian, the probability of SNR to exceed 3 is 0.00135 in the absence of true signals. The probability that at least one out of 80 random realizations exceeds that value is then $1-(1-0.00135)^{80}=0.102$. We find 16 coefficients with SNR$>3$, hence, statistically significant signals seem to be present in more than a dozen VSH terms across the five partitions. This estimation  may be biased toward smaller expected SNR, because the actual distribution of measurement error in astrometry is practically never an ideal Gaussian. Still, if SNR=3 is adopted as a significance threshold, we note that the significant coefficients appear only for five VSH functions (number 1, 2, 3, 5, and 15).

Nine of the formally significant signals are detected for the magnetic terms of first degree (number 1, 2, 3). This unexpected result was also discussed in \citep{2025NatAs...9.1396M}. A portion of this perturbation can be ascribed to the imperfect procedure of Gaia proper motion system adjustment, which was based on a different selection of reference sources using a different algorithm. However, it cannot explain the larger redshift-dependent component of the VSH decomposition. Indeed, the strongest perturbation is seen in the \{mag,0,1,0\} term, which defines a rigid spin around the Galactic poles. The signal changes sign from minus to plus somewhere around $z=1$ or slightly lower, which baffles its interpretation as a systematic error of Gaia astrometry. These estimates are consistent with the previous analysis in \citep[][see Table 4]{2025NatAs...9.1396M}, where the coefficients for the two wider bins of redshifts between 1 and 3 were found to be close to $+13$ \uasyr. The closer CRF objects with $z<1$ were not investigated before, and the different behavior of this VSH term is a new and surprising outcome.

\begin{figure}
    \includegraphics[width=0.95\textwidth]{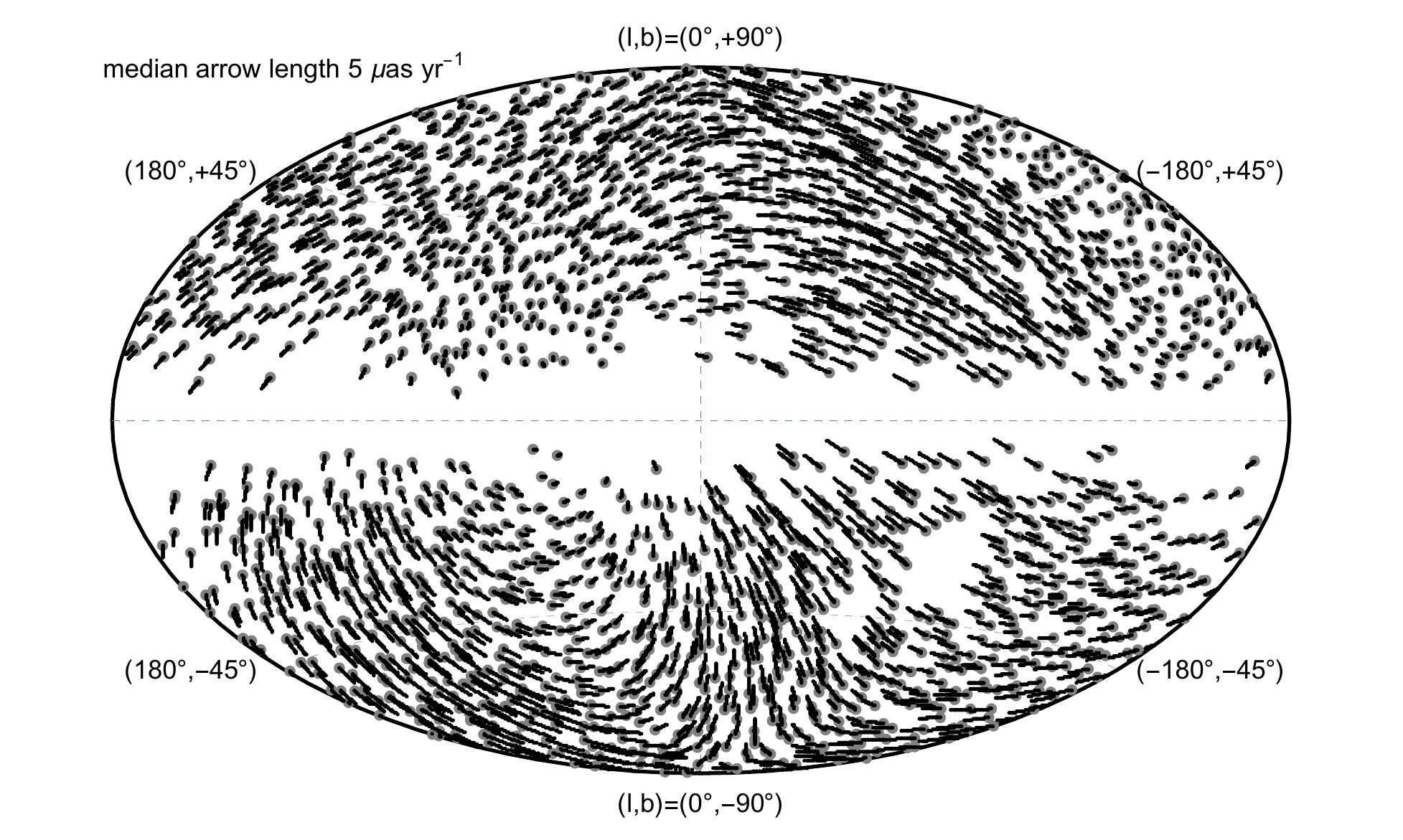}
   \caption{VSH-fitted proper motion field of Gaia CRF quasars with ML-predicted redshifts $0.5<z<1.01$ (Batch 1) on the celestial sphere.\\ Graphical presentation in the Aitoff Galactic projection with the Galactic center direction at the center of the plot. Grey dots at the origin of vectors indicate the mean positions of sources. Only 1\% of sources in this batch are shown. The length and direction of small vectors represent the cumulative fit for the corresponding quasar from the 5 VSH terms with formal SNR$>3$ (Table \ref{res.tab}). The median length of the arrows is 5 \uasyr.}
    \label{field1.fig}
\end{figure}

The \{mag,1,1,1\} term, which defines a rigid spin around the Galactic $X$ axis toward the center of the Galaxy, is found to be significant only in the first batch comprising the nearest CRF sources. It declines to statistically negligible values at $z>1.5$. The remaining rotation term \{mag,2,1,1\} also appears to change sign between the first and second batches (i.e., at approximately redshift 1). It strengthens toward $z\simeq 1.5$ and then starts to decline. The combined pattern of motion defined by the spin harmonics is complex and highly redshift-dependent. We cannot find a cosmological interpretation to these unexpected signals outside models with rotational or torsional components. The remaining possibility is a complex and structured systematic error in the calibration of Gaia proper motion measurements that mimics a redshift dependence. 
This effect could be due to the known magnitude dependent differences between sources in the CRF3 dataset, coming from the calibration of the astrometric instruments \citep{2020A&A...633A...1L,2021A&A...649A...2L}. 
This explanation is further supported by a recent analysis of CRF3 data \citep{2025arXiv250802810T}. 
The detected redshift-dependent spin may be the consequence of hidden correlations with those driving parameters. The statistical relation of variability amplitude to redshift is rather weak at $z>0.5$, however, \citep[Fig. 7 in ][]{2024ApJS..274...27M}, while the distribution of {\tt phot\_bp\_rp\_excess} versus redshift also considerably flattens out for the considered range. Thus, the origin of the spin terms remain an open issue leaving room for a cosmological explanation.

\begin{figure}
    \includegraphics[width=0.95\textwidth]{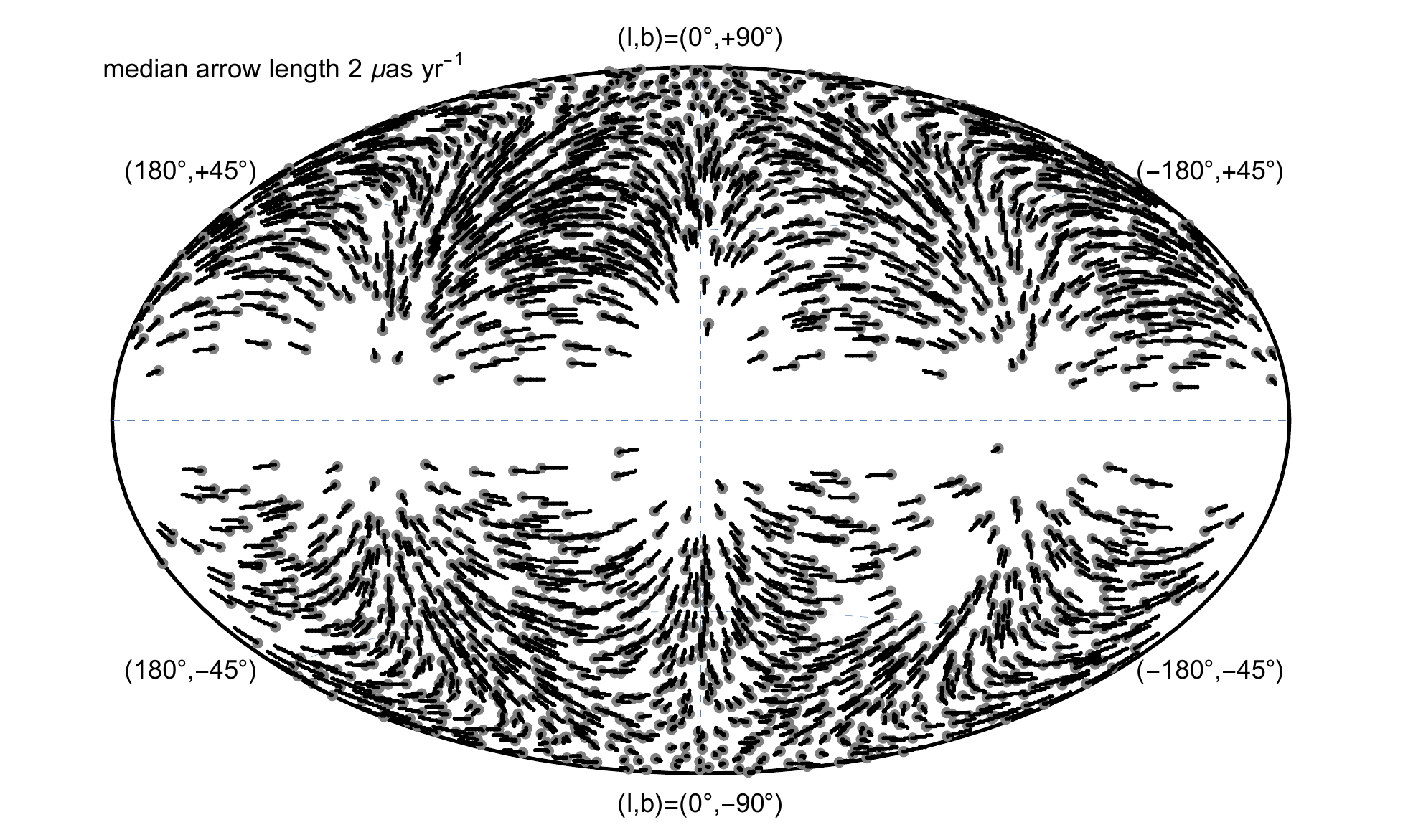}
   \caption{The global vector field representing the quadrupole \{ele,1,2,2\} VSH function (Table \ref{vshlist.tab}), which carries a consistent signal across the range of cosmological redshift. \\ Graphical presentation in the Aitoff Galactic projection with the Galactic center direction at the center of the plot. Grey dots at the origin of vectors indicate the mean positions of quasars in Batch 5. Only 1\% of sources in this batch are shown. The length and direction of small vectors represent the value of the fitted \{ele,1,2,2\} VSH function. The median length of the arrows is 2 \uasyr.}
    \label{ele122.fig}
\end{figure}

Apart from the magnetic first-degree terms, we find only two VSH functions with significant signals. The first-degree \{ele,1,1,1\} function is the only one where the signal is fully expected, because it contains the Galactocentric aberration drift effect, caused by the acceleration of the Solar System relative to the collective rest frame of distant cosmological sources. This determination also serves as a good sanity check. The corresponding amplitudes of the streaming motion for the five batches 1--5 are\footnote{These are obtained from table~\ref{res.tab} as normalized by the coefficients in table~\ref{vshlist.tab}.} $3.58\pm0.69$, $2.01\pm 0.76$, $4.87\pm 0.76$, $4.92\pm 0.80$, and $4.44\pm 0.89$ \uasyr, respectively.
These estimates are systematically lower than the amplitude determined by Gaia from an all-inclusive fit \citep[$5.05\pm 0.35$ \uasyr,][]{2021A&A...649A...9G}, which can be the outcome of the more stringent quality filters applied in this work. We also see a hint at redshift dependence in the determinations for $z<1$, which requires further attention, because aberration is invariant to the redshift of the sources considered. 
The dipole drift may be affected by some high-degree VSH terms that are not included in the fit via the ``harmonic leakage" discussed in \citep{2007AJ....134..367M}. In this case, the zone of avoidance along the Galactic plane (clearly seen in Fgs. \ref{field1.fig} and \ref{ele122.fig}) makes the discretized VSH non-orthogonal, and a sufficiently strong signal in higher degrees can bias the estimates in the lower degrees. 
However, previous detailed analyses of such leakage in CRF3 show robustness within $1 \sigma$ towards corrections of VSH degree $>2$ \cite{2021A&A...649A...9G}, and analyses accounting for such leakage found redshift dependence of a similar strength as in our work \citep{2025arXiv250802810T}.  

The possible leakage of harmonic power from the first-degree magnetic terms (spins) to the electric quadrupole terms, which can include the signal from a Bianchi I cosmology, is a cause for concern. We investigated the full covariance matrices of the VSH coefficients and computed the correlation values from the off-diagonal elements. These values govern the propagation of random uncertainties between the VSH coefficients. Fortunately, most of these $5\times 120=600$ values are smaller than 0.1 in absolute value. We found in total 15 correlations (1 in 40) larger than 0.3 in absolute value. A persistent correlation emerging in all 5 redshift batches is that between terms 3 (VSH \{mag,2,1,1\}) and 11 (VSH \{ele,1,2,1\}) in Table \ref{res.sec} ranging between $+0.33$ and $+0.46$. The fitting results in Table \ref{res.tab} indicate that while the \{mag,2,1,1\} term is large and variable with redshift, the \{ele,1,2,1\} quadrupole is small, negative, and rather stable. Thus, there is no evidence of a significance leakage from term 3 to term 11. The propagation of random error in this case is limited to approximately $0.4\;a^{(M)}_{\{2,1,1\}}\,\sigma_{11}/\sigma_{3}$, which is indeed a small number. 
A negative correlation of $-0.32$ was found for three batches between terms 2 (VSH $\{$mag, 1, 1, 1$\}$) and 12 (VSH $\{$ele, 2, 2, 1$\}$), as well as a persistent correlation of $\sim -0.46$ between terms 6 (VSH $\{$ele, 2, 1, 1$\}$) and 8 (VSH $\{$mag, 1, 2, 1$\}$), which is not relevant to this analysis. 
We refrained from performing a constrained fit omitting the three rotations. If these terms are mostly deterministic rather than stochastic (as defined by hidden systematic errors, for example), which appears to be likely, such a constraint would achieve the opposite effect to the desired outcome by propagating the systematic error into the quadrupole fit.

Fig. \ref{field1.fig} displays the fitted proper motion field for Batch 1 in the Aitoff Galactic projection. Only the five VSH terms with SNR$>3$ are included in the plot. The detected field is therefore formally confident. The pattern reveals that the expected convergence of drift vectors toward the Galactic center (at the center of the plot) is clearly visible only in the southern hemisphere. The north-south asymmetry is coupled with an east-west asymmetry around the principal meridian. The fits in the other batches (not reproduced for brevity) are not too dissimilar, generally representing a complex churning pattern at the achieved level of $\sim 1$ \uasyr. The global field of drift motion appears to include additional signals beside the theoretically predicted aberration dipole.

The only significant quadrupole term detected from this analysis, \{ele,1,2,2\}, is of prime interest in the context of this study. Eq. \ref{mu.eq} shows that the axial anisotropy of a Bianchi I universe is represented by a single \{ele,0,2,0\} harmonic of the global drift in the preferred coordinates $(\hat l,\hat b)$. The coordinate system rotation $(\hat l,\hat b)\rightarrow (l,b)$ from the preferred frame to the standard galactic frame causes the signal to spread between the five electric harmonics of second degree. Therefore, it is possible in principle, that the detected drift in \{ele,1,2,2\} graphically depicted in Fig. \ref{ele122.fig} reflects a general anisotropy in a specific direction. Our task is now to determine if this detection is statistically significant, quantify the corresponding cosmological parameters, and validate these results for the observer's velocity effects.

\section{Interpretation of the velocity field}
\label{inter.sec}
The following consideration of the anisotropic drift is greatly simplified by the axial symmetry of the cosmological model under consideration. Upon coordinate rotation $(\hat l,\hat b)\rightarrow (l,b)$, the single nonzero VSH term \{ele,0,2,0\} transforms into the set of five quadrupole VSH functions of second degree. These rotational transformations of VSH coefficients are defined by Wigner matrices. Specifically, for our case of real-valued functions and coefficients, we have:
\eb 
a^{(E)}_{\{k,2,q\}}=D^{\{0,2,0\}}_{\{k,2,q\}} \hat a^{(E)}_{\{0,2,0\}},
\ee 
where $\hat a^{(E)}_{\{0,2,0\}}$ is the coefficient of the zonal harmonic representing the anisotropy signal in the preferred coordinate frame. The Wigner matrix elements $D$ are functions of the Euler rotation angles $u_1$, $u_2$, and $u_3$ conventionally chosen to be the R3R2R3 sequence. The zonal harmonic \{ele,0,2,0\} is independent of $\hat l$, therefore, the first R3 rotation by $u_1$ is redundant, and the Wigner elements are independent of $u_1$, which can be set to zero. The specific Wigner elements that are consistent with the normalization scheme in Table \ref{vshlist.tab} are defined as:
\begin{eqnarray}
    D^{\{0,2,0\}}_{\{0,2,0\}} &=& \frac{1}{4}+\frac{3}{4} \cos(2 u_2) \nonumber \\
    D^{\{0,2,0\}}_{\{1,2,1\}} &=& \sqrt{\frac{3}{2}}  \sin(2 u_2) \cos u_3 \nonumber \\
    D^{\{0,2,0\}}_{\{2,2,1\}} &=& \sqrt{\frac{3}{2}}  \sin(2 u_2) \sin u_3 \nonumber \\
    D^{\{0,2,0\}}_{\{1,2,2\}} &=& \sqrt{\frac{3}{2}}  \sin^2 u_2 \cos(2 u_3) \nonumber \\
    D^{\{0,2,0\}}_{\{2,2,2\}} &=& \sqrt{\frac{3}{2}}  \sin^2 u_2 \sin(2 u_3) 
    \label{D.eq}
\end{eqnarray}
Within the assumed cosmology, there is a single preferred direction (the $z$-axis), which is independent of redshift. Hence, we have to find a couple of angles $u_2$ and $u_3$ from the entire set of VSH coefficients in Table \ref{res.tab}. To increase the precision of this crucial determination, we can average the coefficients over the five batches for each term. The resulting set of the weighted mean electric quadrupole coefficients (in the same order as in Eqs. \ref{D.eq}) is $(-0.313, -1.870, -0.785, +3.812, +1.668)$ \uasyr\ with formal uncertainties $(0.333, 0.606, 0.572,  0.607, 0.586)$ \uasyr, respectively. Note that only the $a^{(E)}_{\{1,2,2\}}$ coefficient is highly significant, while $a^{(E)}_{\{1,2,1\}}$ and $a^{(E)}_{\{2,2,2\}}$ are marginally significant, and the rest are within the expected random errors of measurement. Using these weighted mean coefficients and their formal errors, we can solve the system of nonlinear equations (\ref{D.eq}) for three unknowns $u_2$, $u_3$, and $s\equiv \hat a^{(E)}_{\{0,2,0\}}$.

Employing the Nelder-Mead optimization method, the following solution was obtained: $s=3.30$ \uasyr, $u_2=1.86$, $u_3=0.22$. We minimized the unweighted sum of squared residuals over the five coefficients for this solution. A solution with ``optimally" weighted residuals (i.e., the $\chi^2$-optimization) was found to produce rather close results. The residuals are smaller in absolute value than their formal uncertainties for all terms except for \{ele,0,2,0\}, where the residual equals $0.93$ \uasyr. The $\chi^2$ over 5 data points was reduced from 59.8 to 8.9 as a result of this fit. To quantify this improvement in terms of statistical confidence, we employ the Fisher ratio distribution paradigm (sometimes called F-test in the literature). The ratio of the sample variances of normalized post-fit residuals and original normalized VSH coefficients is 0.13. The corresponding numbers of degrees of freedom are 2 and 5, since the fit involves 3 fitting parameters. Using the known F-ratio distribution, we directly compute the confidence of the fit at 0.88 as the survival function of this distribution at the given variance ratio. In other words, the probability of the null hypothesis that the original and post-fit variances are equal equals $1-0.88=0.12$. The emerging confidence value can be qualified as marginally significant.

We can now derive the preferred direction of the fitted axisymmetric Bianchi I cosmology as $R_3(u_3)\,R_2(u_2)\,[0,0,1]^T$.
The Galactic coordinates of this estimated axis are $(l_{\rm BI},b_{\rm BI})=(347\degr,17\degr)$.\footnote{Note that the preferred direction is modulo-$\pi$ invariant.} The uncertainties of the derived parameters were evaluated via a Monte Carlo simulation of the nonlinear fit with the VSH coefficients perturbed by normally distributed random values with the given standard deviations. The robust formal error was computed as half the difference between the 0.84- and 0.16-quantiles of the simulated sample. The resulting set of parameters of interest and their formal uncertainties are $s=3.32\pm 0.47$ \uasyr,
$l_{\rm BI}=347.4\degr\pm 4.4\degr$ and $b_{\rm BI}=16.4\degr\pm 5.2\degr$.

The best-fitting amplitude $s(z)$ can now be computed for each redshift batch separately following the same optimization algorithm (but with the fixed rotation angles $u_2$ and $u_3$) from the five VSH coefficients given in Table \ref{res.tab}. We used Monte Carlo simulations with perturbed VSH coefficients and the standard deviations equal to their formal errors. At this step, there is only one free parameter per batch to be fitted, which is the amplitude of the quadrupole zonal harmonic in the rotated frame. The emerging median amplitudes $s(z)$ and their robustly estimated uncertainties are: $3.82\pm 0.78$, $3.13\pm 1.01$, $2.59\pm 0.94$, $3.36\pm 1.01$, $3.53\pm 1.07$ \uasyr\ for batches 1--5, respectively. The corresponding cosmological parameters $\eta'_0+2\bar{\eta}_e$ are $-4.83$, $-3.95$, $-3.26$, $-4.24$, and $-4.46$ \uasyr. The Pearson $\chi^2$ goodness-of-fit test yields a $p$-value of 0.44 for the null hypothesis that the normalized VSH coefficients of this term are drawn from the same normal distribution ${\cal N}[\bar a,1]$. Thus, there is no compelling reason to reject the null hypothesis, and the detected signal is likely to be equal for all the redshift batches. This conclusion is in agreement with the finds in \citep{2025arXiv250802810T}.

This analysis was consistently performed in the galactic coordinate system. In principle, there is a free choice of coordinates, because the total power of VSH terms is invariant to rotation in each degree. For example, one can use the original equatorial system, in which case the detected power of electric quadrupoles would be distributed differently. Our estimation is predicated on the Bianchi I model requirement that the true signal is fully represented by a single (but arbitrarily rotated) zonal quadrupole harmonic and the geometric derivation is sensible only within this model.

\section{Discussion and Future Work} \label{disc.sec}
We have shown in this paper that the simplest uniformly expanding cosmology with anisotropy and a single preferred direction (axisymmetric Bianchi I model) can be testable with the available Gaia data for 1.1 million high-redshift quasars. Specifically, the different cosmic scale in the preferred direction as a function of coordinate time emerges in the global proper motion field as a quadrupole vector spherical harmonic \{ele,0,2,0\}. This field (an example is shown in Fig. 2) is characterized by two opposing poles and an equator plane between them, where the tangential motion nullifies. Depending on the sign of the coefficient, the streaming motion is directed away from the equator toward the poles, or vice versa. Since the coordinate triad of choice (Galactic system in this work) is rotated with respect to the preferred frame, the expected cosmological signal is split unevenly between the 5 electric VSH terms. We performed least-squares VSH fits for five batches of Gaia CRF sources separated by redshifts from 0.5 to 3.0 and determined 16 VSH coefficients along with their formal errors to degree 2. The quadrupole electric terms show a strong signal, which can be the manifestation of a Bianchi I anisotropy.

Using the Wigner matrix functions of VSH rotational transformations and nonlinear optimization, we estimated one single preferred direction that maximizes the signal in the zonal term \{ele,0,2,0\} with a confidence of 0.88. The estimated preferred direction in Galactic coordinates is $(347.4\degr,16.4\degr)\pm (4.4\degr,5.2\degr)$, and the mean amplitude of the signal for the entire collection of sources is $3.32\pm0.47$ \uasyr. The corresponding cosmological drift amplitude $\eta'_0+2\bar{\eta}_e$ is on average $-4.1$ \uasyr\ with statistically insignificant variation across the redshift batches.

Although this Bianchi I fit provides a sizable reduction of the residual $\chi^2$ according to the F-ratio distribution test, there are significant validation challenges for this model. Calculations in Sect. \ref{theo.sec}, Eq. \ref{c.eq}, and Table \ref{coef} indicate that the drift amplitude should be increasing in absolute value due to the greater contribution from the time-variable part $\bar{\eta}_e$. This is seen if we rewrite the quadrupole effect in the proper motion field as:
\eb 
\hat a^{(E)}_{\{0,2,0\}}=-\sqrt{\frac{\pi}{5}}\; \eta'_0\,C(z).
\label{hata.eq}
\ee 
The negative value found in the present analysis implies that $\eta'_0$ is positive, while the dominating term $2\bar{\eta}_e$ is negative and rapidly rising with the lookback time. The universe expands faster in the preferred direction. There are three points of concern for this interpretation:
\begin{enumerate}
    \item Our results are consistent with a flat behavior of $C$ with redshift.
    \item The preferred direction (the axis of the enhanced expansion) is rather close to the Galactic center direction (modulo $\pi$).
    \item There are stronger signals in the magnetic first-degree terms (spins) that are not accounted for in the Bianchi I model.
\end{enumerate}

The isotropic FLRW model predicts a local radial angular velocity field today of $21\,h_0$ \uasyr. For a deviation of $\Delta h_0$ in the preferred direction, the expected proper motion amplitude at $z=0$ amounts to $21\,\Delta h_0/(\sqrt{2} h_0)$ \uasyr. To match our empirical estimate $3.32$ \uasyr, we would have to assume a maximum anisotropy amplitude of $\Delta h_0/h_0=0.22$. Such a high value of the local expansion anisotropy is not confirmed by the currently available analyses. The local expansion rate fluctuation $\Delta H_0$ (related to $\eta'_0$ via $c'/c - a'/a \,=\, 3/2\,\, \eta'\, +\,  O(\eta^2)$)
is estimated in the range 2.4--4.1 km s$^{-1}$ Mpc$^{-1}$ \citep{2023PhRvD.107b3507K}, but most of this directional variation is attributed to the dipole component, while the quadrupole component is smaller in amplitude. The estimated direction of the maximum anisotropy (i.e., the preferred direction in the Bianchi I model) is also discrepant from the local enhancement of $H_0$ \citep{2024A&A...681A..88H}, which is roughly aligned with the CMB dipole \citep{2022PhRvD.105j3510L}. As a caveat, the redshifts investigated in our paper are much higher starting at 0.5 (the astrometry data being relatively unreliable for nearby AGNs). The $C$ ratio defined in Section \ref{theo.sec} is a declining function of redshift reaching $\simeq -28$ at $z=8$ and starting from $-1$ at $z=0$. The latter is a model property by construction. To verify this assumption, we would need to perform the proper motion field analysis for the closest Gaia CRF sources. In lieu of this verification, we can only extrapolate our result to $z=0$.

\begin{figure}
    \includegraphics[width=0.65\textwidth]{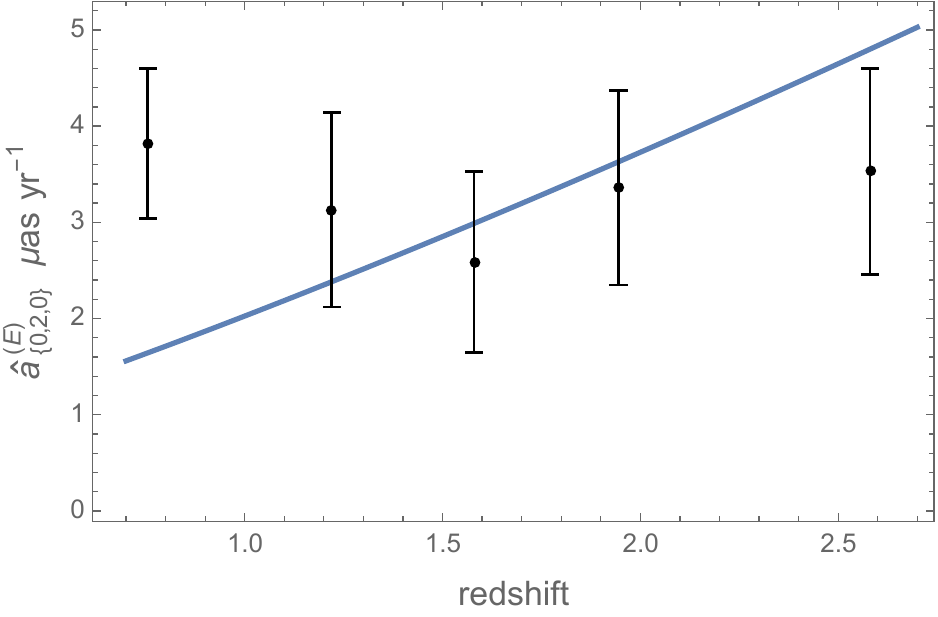}
   \caption{Estimated amplitudes of the quadrupole proper motion harmonic sampled at five redshift batches (black dots with error bars). The semi-empirical Bianchi I model is shown with the solid line. }
    \label{muz.fig}
\end{figure}

Fig. \ref{muz.fig} shows our point-estimates of $s(z)$ values at the centers of the redshift bins with their estimated standard errors, shown as dots with error bars. The solid line shows the theoretical fit following Eq. \ref{hata.eq}. The single fitting parameter here is $\eta'_0$, the time derivative of the scale anisotropy today. It was estimated using the least-squares adjustment of the five estimates of $s(z)$ and the computed corresponding ratios $C(z)$. The fitted value is $\eta'_0=0.76\pm 0.10$ \uasyr. Although the data points do not deviate from the model too far, the general impression is that the model is not successful in describing the empirical results. Formally, the confidence of the model fit can be quantified using the Pearson $\chi^2$-test on the post-fit error-normalized residuals ``data$-$model". The resulting $p$-value of 0.44 is inconclusive, in that the null hypothesis that the post-fit residuals are normally distributed with the given standard deviations should not be rejected. There is no compelling evidence either that the data conforms to the model.

According to the general theorem by \citet{1983PhRvD..28.2118W} (almost) all Bianchi universes with positive cosmological constant and mild constraints on the energy momentum tensor have an exponential decrease of the anisotropy with growing cosmic time. Therefore, the anisotropy increases with growing redshift as shown in Fig. \ref{muz.fig}. 

The uncertainties and the error bars of the previously available data have been too large to rule out (axial) Bianchi I by its fits to the CMB ($z \sim 1100$) 
\citep{2014MNRAS.441.1646C} 
and to the Lema{\^i}tre-Hubble diagram ($z < 1.3$) 
\citep{Schucker:2014wca}
,  although the estimated Hubble stretch $ \eta'_0/H_{F0}$ from the CMB fit was found to be $10^{-8}$ times smaller and of opposite sign with respect to the Hubble stretch from the latter fit, see Table \ref{compar}. 
However the discrepancy between the CMB fit and the present Gaia fit is significant.

\begin{table}[!h]
\begin{center}
\caption{Comparison with two other fits}
 \label{compar}
 \vspace{5mm}
\begin{tabular}{|c|c|c|c|c|c|}
\hline
&&&&\\[-4mm]
data&redshift &$\eta'_0/H_0$&galact. long.
&galact. lat.&reference
\\[2mm]\hline\hline
SuperNovae 1a&$z<1.3$&$-(1.3\pm0.9)\%$&$106^\circ\pm16^\circ$&
$1^\circ\pm10^\circ $&S,T\&V (2014)
\\\hline
Gaia&$z<3$& ${}\hspace{3mm} (5.0\pm0.7)\%$&
$347.4^\circ\pm 4.4^\circ$&$16.4^\circ\pm5.2^\circ$
&ibidem
\\\hline
CMB&$z\sim1100$&$(3.8\pm1.2)\cdot10^{-8}$&$-$
&$\pm17^\circ$&Cea (2014)
\\\hline
\end{tabular}
 \end{center}
\end{table}

In the Lema{\^i}tre-Hubble diagram, axial Bianchi I universes are characterized  by a scalar quadrupole, whose amplitude also increases with redshift. 
Such a signal was found by \citet{Sorrenti:2024ztg}  in the Pantheon+SH0ES data. 

Still, together with the results of this analysis, a Bianchi I model does not emerge as a high-priority paradigm for describing the time-evolution of the Universe. On the technical side, the weak condition of our method is rooted in the level of precision that has been achieved thus far ($\sim 1$ \uasyr). The prospects for the direct kinematic mapping method are quite good, however. In a year, the next data release of Gaia is expected to improve the precision of proper motions by a factor 2--5. In a longer time frame, a second epoch, second generation Gaia-like space mission with infrared capabilities will push the sensitivity to well below $0.1$ \uasyr\ and obtain a greater number of high-redshift sources \citep{2021ExA....51..783H}. A deeper, more critical understanding of Gaia systematic errors becomes the crux of the problem.
Prelaunch analysis of error propagation in Gaia measurements has been limited to approximations of the single object covariance largely ignoring the correlated components due to calibration errors, for example \citep{2012A&A...543A..14H}. The limitations are caused by the sheer size of the adjustment problem, which would require inversion of a matrix of $\sim10^{20}$ elements. Furthermore, the errors of VSH coefficients can be both of random and deterministic origin. Systematic errors of Gaia proper motions thus remain to be better understood. In lieu of a thorough error propagation model, secondary characterization techniques should be explored, such as mapping explicit correlations between various primary and metadata parameters.

The dipole electric terms in Eq. \ref{mu.eq} reflect the global proper motion field arising from the peculiar velocity of the observer in the comoving cosmic frame. These effects are too small to be detected with the present-day astrometric precision. The corresponding coefficients in Table \ref{res.tab} are indeed consistent with zero with the notable exception of $a^{(E)}_{\{1,1,1\}}$, which shows a consistent signal. The origin of the signal is in the apparent dipole drift of all celestial objects (mostly) toward the Galactic center caused by the acceleration of the Solar system barycenter and the resulting rotation of the secular aberration dipole. This expected relativistic effect is reproduced in our analysis. However, the variations of this coefficient between the redshift batches is somewhat higher than expected, since the aberration amplitude is redshift-independent. Furthermore, the weighted mean amplitude of the proper motion dipole is $3.96\pm 0.65$ \uasyr\ instead of the previously estimated $5.1\pm 0.4$ \uasyr\ \citep{2021A&A...649A...9G}. The result for batch 2 (Table \ref{res.tab}) appears to be especially deviant on the low side. We tentatively interpret this difference as the consequence of the data quality filters (Sect. \ref{data.sec}). In particular, we discarded all Gaia data with 6-parameter solutions, which were found to produce deviant VSH fits from a dedicated analysis of systematic errors.

The presence of large and confidently detected magnetic harmonics of first degree (coupled with the absence of such signals in the 2nd degree magnetic terms) poses a challenge to the considered Bianchi I model, which allows only radial expansion of spacetime. We should note that the sample-mean values of these spin harmonics are nonessential because of the intrinsic degeneracy of global astrometric solutions. Astrometric facilities (Gaia and VLBI) measure only arcs between sources, which are invariant with respect to rotation and spin of the entire system. The surprising result is the apparent strong dependence of the corresponding coefficients presented in Table \ref{res.tab} on redshift. All the three components (number 1,2,3) appear to change sign around $z=1$. The origin of these signals is unclear. Apart from the mundane (but not easily verified or evidently self-consistent) explanation that these detections stem from hidden systematic errors of Gaia proper motions, one could consider more complex G{\"o}del-like cosmologies with rotation that may be redshift-dependent. A slow rotation of the Universe around the axis of the anisotropy dipole was shown to have the potential of resolving the $H_0$ tension \citep{2025MNRAS.538.3038S} but the upper limit on the angular rate today (constrained from assuming subluminous tangential velocities at the Hubble horizon) is 0.4 \uasyr\, which is an order of magnitude smaller than our estimates for larger redshifts. However, as the spin rate is expected to increase with lookback time proportionally to $a^{-2}(t)$, a cosmological interpretation of the unexpected magnetic harmonics deserves further careful examination. 

\section*{Acknowledgments} 
AH acknowledges the Perren Fund at the University of London and the Astronomy Unit at Queen Mary University of London for their support. 
\bibliography{main}{}
\bibliographystyle{aasjournal}

\section{Appendix}

For the fitting of theoretical redshift and drift to observations, we need the  
comoving geodesic distance $\chi_{e0}$. Rather than relying on numerical integrals, we use analytic expressions due to \cite{1972MNRAS.159...51E}
 and follow the notations of 
\citet{Valent:2022smp,Valent:2023tos}. 

We need two fixed numbers:
\begin{eqnarray}
k^2:=1/2+\sqrt{3}/4=0.933301\cdots,
\quad
q:=\exp(-\pi K'/K)=0.163034\cdots,
\end{eqnarray}
with the elliptic integral,
\begin{eqnarray}
K(k^2):=\int_0^{\pi/2}\frac{d x}{\sqrt{(1-x^2)(1-k^2\,x^2)}}, 
\quad
k^2\in[0,1],
\end{eqnarray}
 and $K'(k^2):=K(1-k^2)$. (The prime does not mean a derivative.) In the definition of $q$ the elliptic integrals are evaluated at our fixed value of $k^2$.
 
We need the Jacobi elliptic function cn\,$(u,k^2)$, with $u\in \mathbb{R}$  and $k^2\in[0,1]$. Since we keep $k^2$ constant, we will not write  the second variable and denote by ${\rm arc}\,{\rm cn}\,(u)$ the inverse of cn\,$(u)$. Similar definitions hold for ${\rm sn}\,(u)$ and ${\rm dn}\,(u)$.

We also need the first of the four Jacobi $\theta $-functions, 
\begin{eqnarray}
\theta _1(v,q):=2\sum_{n\geq 0}(-1)^n\,q^{(n+1/2)^2}\,\sin((2n+1)v),
\quad
v\in \mathbb{R}, \quad q\ \text{as above},
\end{eqnarray}
and define 
\begin{eqnarray}
 H(u):=\theta _1\left(\frac{\pi u}{2K},q\right),
\quad \text{and}\quad 
\zeta(u):=\,\frac{d}{d u}\, \ln H(u).  
\end{eqnarray}
Table \ref{access} summarizes a few commands accessing Jacobi elliptic functions and others in Mathematica and Maple.
\begin{table}[!h]
\begin{center}
\caption{A few access commands in Mathematica and Maple. Note the absence of the square on $k$ in Maple. Note also that EllipticTheta[$i$, v, q]  are not  elliptic  functions.}
\begin{tabular}{|c||c|c|}
\hline
& Mathematica&Maple\\\hline\hline
$K(k^2)$&EllipticK[k$\,\hat{}\,$2]&EllipticK(k)\\\hline
${\rm cn}\,(u,k^2)$&JacobiCN[u,k$\,\hat{}\,$2]&JacobiCN(u,k)\\\hline
arc$\,{\rm cn}\,(u,k^2)$&InverseJacobiCN[u,k$\,\hat{}\,$2]&InverseJacobiCN(u,k)\  \\\hline
$\theta _i(v,q),\ i=1,2,3,4$&EllipticTheta[$i$, v, q]&JacobiTheta$i$(v,q)\\\hline
$H_2F_1(a,b;c;x)$&
Hypergeometric2F1[a, b, c, x]&hypergeom([a,b],[c],x)
\\\hline
\end{tabular}
 \label{access}
 \end{center}
\end{table}

Let us define the conformal time $\chi $ as a function of cosmic time $t$ by
\begin{eqnarray}
\chi (t)\, : =\int_0^t \,\frac{d \tilde t}{a_F(\tilde t)}\,.
\end{eqnarray}
The big bang happens at $t=0$, $a_F(0)=0$, but the singularity there is integrable, $\chi (0)=0$. As cosmic time tends to infinity, the dimensionless conformal time remains finite, $\chi \in [0, \widehat\chi ] $ with 
\begin{eqnarray}
\widehat\chi : =\int_0^\infty \,\frac{d \tilde t}{a_F(\tilde t)}&=&
4\sqrt{\pi }\,\frac{\Gamma (7/6)}{\Gamma (2/3)}
\,\frac{1}{R}\, 
\frac{1}{a_{F0}\,\sqrt{\Lambda} },
\label{chic}
\\[2mm]
R:=\left(\frac{1-\Omega _{\Lambda 0}}{\Omega _{\Lambda 0}}\right)^{1/3}, &&4\sqrt{\pi }\,\frac{\Gamma (7/6)}{\Gamma (2/3)}\,=4.8573\cdots.
\end{eqnarray}

The central formula of this appendix is:
\begin{eqnarray}
t(\chi )=\sqrt{\frac{3}{\Lambda }}\left\{ \!{\textstyle\frac{1}{2}} \ln\!\!\left(\!
\frac{H(\sigma (\chi +\widehat\chi))}{H(\sigma (-\chi +\widehat\chi))}
\,\frac{\widehat{s}\,\,{\rm dn}\,(\sigma \chi )+\widehat{d}\,\, {\rm sn}\,(\sigma \chi )}{\widehat{s}\,\,{\rm dn}\,(\sigma \chi )-\widehat{d}\,\,{\rm sn}\,(\sigma \chi )\!}\right)
\!\!-\!\!\left( \!\!\zeta(\sigma \widehat\chi)   +\frac{\widehat{d}}{\widehat{s}}\,\!\right)\!\sigma \chi \right\} ,
\nonumber
\end{eqnarray}
\vspace{-6mm}
\begin{eqnarray}
\label{rr2}
\end{eqnarray}
where we use the abbreviations:
\begin{eqnarray}
\sigma :=3^{-1/4}R\,a_{F0}\,\sqrt{\Lambda},
\hspace{7mm}
&&
\widehat{s}:={\rm sn}\,(\sigma\widehat\chi)=\,\frac{2\cdot 3^{1/4}}{1+\sqrt{3}},
 \\[2mm]
\widehat{c}:={\rm cn}\,(\sigma\widehat\chi)=\,\frac{ 1-\sqrt{3}}{1+\sqrt{3}},&\quad&
\widehat{d}:={\rm dn}\,(\sigma\widehat\chi)=\,\frac{1}{1+\sqrt{3}}\,.
\label{hat}
\end{eqnarray}
Note that the first of equations (\ref{hat}) can be written
\begin{eqnarray}
\widehat\chi=\,\frac{1}{\sigma}\, {\rm arc}\,{\rm cn}\,\!\!\left(\frac{ 1-\sqrt{3}}{1+\sqrt{3}}\right),
\quad\quad
3^{1/4}\,{\rm arc}\,{\rm cn}\,\!\!\left(\frac{ 1-\sqrt{3}}{1+\sqrt{3}}\right)=\,4.8573\cdots,
\end{eqnarray}
which coincides with equation (\ref{chic}). 

The central formula (\ref{rr2})  allows us to rewrite the scale factor $a_F(t)$ as a function of conformal time:
\begin{eqnarray}
a_F(t(\chi ))\,=\,a_{F0}\,\frac{R}{1+\sqrt{3}} \,\frac{1-{\rm cn}\,(\sigma \chi )}{{\rm cn}\,(\sigma \chi)-{\rm cn}\,(\sigma \widehat\chi)}\,.   \label{rr1}
\end{eqnarray}
From this we deduce the age of the universe today in dimensionless conformal time,
\begin{eqnarray}
\chi_0:=\chi(t_0)=\,\frac{1}{\sigma}\, {\rm arc}\,{\rm cn}\,\!\!\left(\frac{R+ (1-\sqrt{3})}{R+(1+\sqrt{3})}\right),
\label{conformalAge}
\end{eqnarray}
and the conformal emission time of a photon arriving today with a redshift $z$,
\begin{eqnarray}
\chi_e:=\chi(t_e)=\,\frac{1}{\sigma}\, {\rm arc}\,{\rm cn}\,\!\!\left(\frac{R+ (1-\sqrt{3})/(z+1)}{R+(1+\sqrt{3})/(z+1)}\right).
\label{conformalEmTime}
\end{eqnarray}
Finally the comoving geodesic distance between the source and the observer or the conformal time of flight is $\chi _{e0}=\chi _0-\chi _e$.

An alternative formula for $\chi_{e0}$  derives from Eq. \ref{chi.eq} and Eq. \ref{scale}:
\begin{eqnarray}
\chi_{e0}= \frac{1}{a_{F0}}\sqrt{\frac{6}{\Lambda}}\left({\cal U}(t_0)-\left(\frac{\cosh{(\sqrt{3 \Lambda} t_0)}-1}{\cosh{(\sqrt{3 \Lambda} t_e)}-1}\right)^{1/3}{\cal U}(t_e)\right),
\label{uu}
\end{eqnarray}
where
\begin{eqnarray} 
{\cal U}(t):= \left(\cosh(\sqrt{3 \Lambda}\;t)+1\right)^{1/2}\,{\rm H_2F_1}\left(\frac{1}{6},\frac{1}{2},\frac{7}{6},-\sinh^2(\sqrt{3 \Lambda}\;t/2)\right)\tanh(\sqrt{3 \Lambda}\;t/2),\nonumber\\
{}
\end{eqnarray}
and $H_2F_1$ is the hypergeometric $_2F_1$ function.

\end{document}